\documentclass[aps,prl,superscriptaddress,showpacs,floatfix,nofootinbib,notitlepage,twocolumn]{revtex4-1}
\usepackage{amsmath,graphicx,float,hyperref,colortbl,upgreek}

\begin{document}

\title{Dilepton polarization as a signature of plasma anisotropy}

\author{Maurice Coquet}
  \email{maurice.louis.coquet@cern.ch}
\author{Michael Winn}
 \email{michael.winn@cern.ch}
\affiliation{
Universit\'e Paris-Saclay, Centre d'Etudes de Saclay (CEA), IRFU\\ D\'epartement de Physique Nucl\'eaire (DPhN), Saclay, France
}

\author{Xiaojian Du}
 \email{xiaojian.du@usc.es}
\affiliation{
Instituto Galego de Fisica de Altas Enerxias (IGFAE), Universidade de Santiago de Compostela, E-15782 Galicia, Spain
}

\author{Jean-Yves Ollitrault}
 \email{jean-yves.ollitrault@ipht.fr}
\affiliation{
 Universit\'e Paris-Saclay, CNRS, CEA, Institut de physique th\'eorique, 91191 Gif-sur-Yvette, France
}

\author{Sören Schlichting}
 \email{sschlichting@physik.uni-bielefeld.de}
\affiliation{
 Fakult\"at f\"ur Physik, Universit\"at Bielefeld, D-33615 Bielefeld, Germany
}

\date{\today}

\begin{abstract}
We propose the angular distribution of lepton pairs produced in ultrarelativistic heavy-ion collisions as a probe of thermalization of the quark-gluon plasma. 
We focus on dileptons with invariant masses large enough that they are produced through quark-antiquark annihilation in the early stages of the collision. 
The angular distribution of the lepton in the rest frame of the pair then reflects the angular distribution of quark momenta. 
At early times, the transverse pressure of the quark-gluon plasma is larger than its longitudinal pressure as a result of the fast longitudinal expansion, which results in an oblate lepton distribution. 
By contrast, direct (Drell-Yan) production by quarks and antiquarks from incoming nuclei, whose momenta are essentially longitudinal, results in a prolate distribution. 
As the invariant mass increases, Drell-Yan gradually becomes the dominant source of dilepton production, and the lepton distribution evolves from oblate to prolate. 
The invariant mass at which the transition occurs is highly sensitive to the equilibration time of the quark-gluon plasma or, equivalently, the shear viscosity over entropy ratio $\eta/s$ in the early stages of the collision. 
\end{abstract}

\maketitle

{\it Introduction. ---} 
Ultrarelativistic heavy-ion collisions produce a quark-gluon plasma which gradually thermalizes as it expands into the vacuum~\cite{Busza:2018rrf}. 
There is wide consensus on the theory side~\cite{Schlichting:2019abc,Berges:2020fwq} that before it approaches local thermal equilibrium, there is a pre-equilibrium phase during which the longitudinal pressure is smaller than the transverse pressure. 
The mechanism generating this asymmetry is the rapid longitudinal expansion of the fluid at early times~\cite{Bjorken:1982qr}. 
It is well known from cosmology that isotropic expansion generates a redshift, i.e., that particle momenta decrease with time in the comoving rest frame of the fluid. 
If the expansion is mostly longitudinal, this redshift affects primarily the longitudinal momentum.  
The pressure anisotropy is the standard quantity which characterizes the deviation from local equilibrium~\cite{Baier:2000sb,Epelbaum:2013ekf,Berges:2013fga,Kurkela:2015qoa,Kurkela:2018vqr}, but its experimental signatures have remained elusive so far~\cite{Alqahtani:2017mhy}. 
We propose to detect it by making use of the kinematic properties of electron-positron or muon-antimuon pairs from the collision, known as dileptons. 

So far, these properties have been discussed in the context of dilepton production by a thermalized quark-gluon-plasma~\cite{Hoyer:1986pp} or by hadronic sources~\cite{Bratkovskaya:1995kh}~\footnote{Photon polarization has been found to be sensitive to the plasma anisotropy~\cite{Ipp:2007ng, Baym:2014qfa}, a more challenging objective than the reconstruction from the dilepton kinematics.}. For the first time, we point out the direct and model-independent connection between intermediate mass dilepton kinematics and pressure anisotropy at early times, based on explicit calculations. 

Dileptons are produced throughout the collision, and the vast majority reach the detectors without undergoing further interactions, so that they are golden signatures of the early history, unlike hadrons. 
The invariant mass $M$ of a dilepton gives a handle on its production time. 
On average, the larger $M$, the earlier it is produced. 
The chronology is as follows. 
First comes production through the annihilation of a quark and antiquark belonging to incoming nuclei, known as the Drell-Yan process~\cite{Drell:1970wh}, which is the dominant dilepton source for the largest values of $M$, apart from narrow resonances (e.g., bottomonium decays).  
Then follows quark-antiquark annihilation in the quark-gluon plasma, which itself starts with pre-equilibrium emission~\cite{McLerran:1984ay}. 
This source, which is the focus of our study,  is expected to exceed Drell-Yan production  for masses $M\lesssim$3-4~GeV~\cite{Coquet:2021lca}, depending on how fast the quark-gluon plasma thermalizes. 
Later emission in the hadronic phase is expected to be significant only for much lower masses, $M\lesssim 1.2$~GeV~\cite{Rapp:2013nxa,Song:2018xca}. 
Therefore, our main focus is the intermediate mass region $1.2<M<5$~GeV, excluding the narrow resonances from charmonium decays~\cite{Andronic:2015wma}.

%Dilepton production in heavy-ion collisions is among the most difficult observables. 
Isolating these dileptons experimentally is very challenging. 
%At any collision energy, the identification of primary leptons against the dominant production of hadrons, mostly pions, and the combinatorial background of dileptons from uncorrelated sources poses a challenge. 
The measured leptons include the primary leptons of interest, and also secondary products of various processes (hadronic decays, and photon conversions in the case of electrons). 
Uncorrelated lepton pairs result in a combinatorial background, whose relative magnitude increases with the system size and is therefore large in heavy-ion collisions. 
This uncorrelated background is typically subtracted by mixing leptons from different events. 
%In addition, the correlated dilepton production  is dominated by the large background created by weak semileptonic decays of hadrons containing a charm quark at LHC energies in the intermediate mass region~\cite{ALICE:2018ael,ALICE:2023jef}, where the largest sensitivity to the early stages of the collision can be achieved~\cite{Coquet:2021lca}. 
For dileptons in the intermediate mass range, which constitute the focus of our study, there is in addition a {\it correlated\/} background from weak decays of charmed hadrons, which always come from charm-anticharm pairs.  
At LHC energies, this background is large enough~\cite{ALICE:2018ael,ALICE:2023jef} that it has prevented the observation of dilepton emission from the plasma so far. 
%Hence, the emission of dileptons from the plasma in the intermediate mass region has not yet been separated experimentally from the leptons from these weak decays.  
%However, background and signal separation is in principle possible~\cite{ALICE:2023jef}. 
In principle, however, it can be isolated by measuring the slight displacement, by a fraction of a millimeter (or in the millimeter range at forward rapidity, thanks to the Lorentz boost), of the vertex of the weak decay relative to the primary vertex of the collision~\cite{ALICE:2023jef}. 
%In addition, the upgrades of ALICE and LHCb that took large luminosity heavy-ion data for the first time in 2023 will provide improved detector performances that should allow first yield measurements. 
%In fact, this measurement is one of the key motivations for the ALICE upgrade. 
The upgrades of ALICE (largely motivated by this measurement) and LHCb, which took high-luminosity heavy-ion data for the first time in 2023, provide improved detector performances that should allow first yield measurements. 
%Furthermore, with the future detector projects LHCb Upgrade 2~\cite{LHCb:2018roe} and ALICE 3~\cite{Adamova:2019vkf} with further performance improvements, a sufficient rejection of the charm background in most central collisions should be within reach, by measuring the slight displacement, by a fraction of a millimeter or in the millimeter range for large Lorentz boosts at forward rapidity, of the vertex of the weak decay relative to the primary vertex of the collision. 
Further improvements are expected along with the detector projects LHCb Upgrade 2~\cite{LHCb:2018roe} and ALICE 3~\cite{Adamova:2019vkf}, and a sufficient rejection of the charm background should be within reach in a decade. 

\begin{figure}
    \centering
    \includegraphics[width=.5\textwidth]{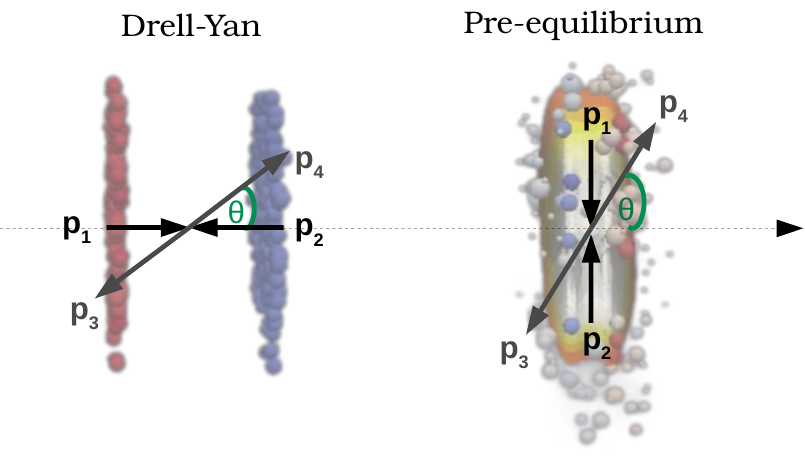}
    \caption{Illustration of the typical kinematic configuration of leading-order dilepton production from the Drell-Yan process (left) and from the quark-gluon plasma in the pre-equilibrium stage (right). The dotted arrow represents the longitudinal $z$-axis, parallel to the beam line. 
    The polar angle in the Collins-Soper frame is labeled $\theta$.
    }
    \label{fig:cartoon}
\end{figure}
The kinematics of lepton pair production can be used to probe the anisotropy of the quark momentum distribution, which itself reflects the pressure anisotropy in the quark-gluon plasma~\cite{Baym:2017qxy}. 
The leading-order process is $q\bar q \to l^+l^-$, where $l^-$ and $l^+$ denote the lepton and its antiparticle. 
In the center-of-mass frame, if one neglects the lepton mass (which is an excellent approximation for both muons and electrons in the considered invariant mass  range), the distribution of leptons per solid angle is proportional to $1+\cos^2\theta$, where $\theta$ denotes the angle between quark and lepton momenta.\footnote{This is due to the fact that the interaction is mediated by a spin-1 particle, hence the term of ``polarization'' traditionally used to characterize this effect~\cite{Speranza:2018osi}, even though it refers to the unpolarized cross section.}
Therefore, emission parallel to the quark is more probable by a factor two than perpendicular to the quark. 
The Drell-Yan process, where quark momenta are mostly longitudinal, will therefore result in preferential emission of longitudinal leptons, while pre-equilibrium emission in the quark-gluon plasma, where quark momenta are mostly transverse in the rest frame of the fluid, will result in preferential emission of transverse leptons. 
This is illustrated schematically in Fig.~\ref{fig:cartoon}.

The natural observable to quantify this effect is the distribution of the angle between the positive lepton and the $z$-axis in the rest frame of the dilepton. This axis corresponds to the beam direction\footnote{In general, the two colliding beams are not exactly collinear, and the "beam direction" is defined as the bisector between the two beams.}.
The defined angle is referred to as the polar angle of the Collins-Soper reference frame~\cite{Collins:1977iv}, which we denote by $\theta$. Neglecting the lepton masses, the cosine of this angle is computed from the momenta ${\bf p}_3$ and ${\bf p}_4$ and energies $E_{3}=p_3$ and $E_{4}=p_4$ of $l^+$ and $l^-$  in the laboratory frame using the standard formula: 
\begin{equation}
\label{CS}
\cos \theta=\left.\frac{ p_{4}^{z}}{|{\bf p}_{4}|}\right|_{\text{CS frame}}=\frac{2\left(E_3 p_4^z - E_4 p_3^z\right)}{M \sqrt{M^2+p_T^2}},
\end{equation}
where  $p_T$ in the denominator denotes the transverse momentum of the dilepton, $p_T\equiv\left| {\bf p}_{T3}+{\bf p}_{T4}\right|$. This angular distribution has been measured in fixed-target 158A GeV In-In collisions by NA60~\cite{NA60:2008iqj} demonstrating the principle measurement feasibility. At these collision energies, the space-time picture used here is however not applicable. 

We first evaluate the distribution of $\cos\theta$ for dileptons emitted by the quark-gluon plasma,  including the pre-equilibrium stage. 
This calculation is done along the lines of our previous works~\cite{Coquet:2021gms, Coquet:2021lca}.
We then carry out a similar calculation for the  Drell-Yan process. 
We finally present our results for the sum  of these two contributions, and show how it evolves as a function of the invariant mass $M$. 

\bigskip
{\it Dilepton emission by the quark-gluon plasma. ---} 
We calculate the production rate of dileptons to leading order in perturbation theory, that is, the rate of quark-anti-quark annihilation, without any additional gluon in the initial or final stage. 
We denote by  $f_q (\mathbf{p}_1,x)$ and $f_{\Bar{q}} (\mathbf{p}_2,x)$ the phase-space  distributions of quarks and anti-quarks, where ${\bf p}_1$ and ${\bf p}_2$ are their momenta, and $x$ denotes the space-time coordinate where annihilation occurs.  
The production rate of a lepton pair with momenta ${\bf p}_3$ and ${\bf p}_4$ is obtained by summing the cross section over ${\bf p}_1$ and ${\bf p}_2$ : 
\begin{align}
\frac{dN}{d^4xd^3{\bf p}_3d^3{\bf p}_4} & = \frac{e^4}{M^4}\frac{\sum q_f^2}{(2\pi)^3 (2p_3)(2\pi)^3 (2 p_4)}\nonumber \\
& \times \int  \frac{d^3{\bf p}_1}{(2\pi)^3 (2p_1)} \frac{d^3{\bf p}_2}{(2\pi)^3 (2 p_2)} f_{q}({\bf p}_1) f_{\bar{q}}({\bf p}_2) \nonumber \\
& \times l_{\mu\nu}\Pi^{\mu\nu} (2\pi)^{4} \delta^{(4)}\left(p_1+p_2-p_3-p_4\right), 
\end{align}
where the squared matrix element $l_{\mu\nu}\Pi^{\mu\nu}$ is given by~\cite{Schwartz:2014sze}:
\begin{eqnarray}
\label{matrix}
    l_{\mu\nu}\Pi^{\mu\nu} & = & 32N_c\left[ (p_1.p_3)(p_2.p_4) + (p_1.p_4)(p_2.p_3) 
%    + (p_1.p_2)m^2
\right].
\end{eqnarray}
In the rest frame of the dilepton, where ${\bf p}_2=-{\bf p}_1$ and ${\bf p}_4=-{\bf p}_3$, Eq.~(\ref{matrix}) yields the usual $1+\cos^2\theta$ distribution, where $\theta$ is the angle between ${\bf p}_4$ and ${\bf p}_2$. 

The next step is to model the phase-space distributions $f_{q,\bar q}({\bf p},x)$ of quarks and antiquarks. 
The calculation closely follows the steps of our previous works~\cite{Coquet:2021gms, Coquet:2021lca}, and we only recall the main points, referring to these earlier works for technical details. 

We first discuss the momentum dependence. 
In thermal equilibrium, the phase-space distribution would be a Fermi-Dirac distribution at temperature $T(x)$ in the local rest frame. 
In order to describe the pre-equilibrium stage, two modifications must be brought to this equilibrium distribution.  

The first modification is the momentum anisotropy, which implies that the quark distributions favor large transverse momenta over longitudinal momenta. 
We encapsulate this property by making the following replacement in the expression for the quark/anti-quark distributions: $|\mathbf{p}| \rightarrow \sqrt{p_t^2+\xi^2 p_z^2}$, where $\xi>1$ is an anisotropy parameter~\cite{Romatschke:2003ms,Martinez:2007pjh}. 
Our main goal in this paper is to see this momentum anisotropy through the produced leptons.

The second modification that one must include is the underpopulation of quarks with respect to the equilibrium value. 
Indeed, the initial state of heavy-ion collisions is mainly comprised of gluons, which dominate the phase space~\cite{Lappi:2006fp}. 
The production of quarks and the approach towards chemical equilibrium only develops incrementally through partonic interactions~\cite{Kurkela:2018oqw, Kurkela:2018xxd}. 
This effect is taken into account by multiplying the momentum distributions by a global ``quark suppression'' factor $q_s$, which is smaller than unity.

We now explain how the space-time dependence is modeled. 
We assume for simplicity that the quark-gluon plasma undergoes uniform longitudinal expansion~\cite{Bjorken:1982qr}. 
We neglect transverse flow, which develops linearly as a function of time~\cite{Vredevoogd:2008id}, and is therefore negligible at the early times which we consider. 
This is likely to be a poor approximation for the lowest invariant masses considered in this paper,  but the effect of transverse flow on dilepton polarization for these low masses has already been studied elsewhere~\cite{Speranza:2018osi}. 
In addition, we assume that the transverse density is uniform. 
We have shown previously that taking into account the inhomogeneity of the density profile amounts to a modest overall rescaling~\cite{Coquet:2021gms}. 
Thanks to these simplifications, the integral of the production rate over space-time coordinates reduces to an integral over time~\cite{Coquet:2021gms}. 

Both the anisotropy parameter $\xi$ and the quark suppression factor $q_s$ are computed as functions of time using QCD kinetic theory~\cite{Giacalone:2019ldn, Du:2020dvp, Du:2020zqg}. 
Even though this is, strictly speaking, a weak-coupling approach, it has been well established over the course of the last few years that the physics of isotropization is  
similar in the strong-coupling limit, the only difference being the time scale of the isotropization process~\cite{Giacalone:2019ldn}. 
The control parameter which determines this time scale is the ratio of the shear viscosity over entropy of the quark-gluon plasma, $\eta/s$~\cite{Kovtun:2004de,Romatschke:2007mq,Heinz:2013th}, which is assumed to be constant in our calculation,  of which it is the only free parameter. 

QCD kinetic theory provides the evolution of pressure anisotropy and of the fraction of energy density carried by quarks as a function of time.
Our calculation is initialized at $\tau=0.01$~fm/c and ends at $\tau=20$~fm/c, a time range large enough that it encompasses essentially all dilepton production. The default initial conditions of our calculation are $\xi\gg 1$ and $q_s=0$, that is, the momenta in the initial state are purely transverse, and there are only gluons in the initial state. 
We neglect the quarks which are present initially from the incoming nuclei. 
In this respect, our calculation underestimates dilepton production in the quark-gluon plasma. 
In order to assess the effect of quark suppression, we will also show results in the other extreme where quarks are in chemical equilibrium at all times, that is, $q_s=1$.

%The control parameter which determines how fast the pressure isotropizes is the ratio of the shear viscosity over entropy of the quark-gluon plasma, $\eta/s$~\cite{Kovtun:2004de,Romatschke:2007mq,Heinz:2013th}, which is assumed to be constant in our calculation, of which it is the only free parameter.
Dimensional analysis shows that the pressure anisotropy $\xi$ and the quark suppression factor $q_s$ solely depends on the dimensionless variable $\tilde{w}=\tau T /(4 \pi \eta / s)$~\cite{Heller:2016rtz}, where $\tau$ is the time in the rest frame of the fluid, and $T$ the temperature at time $\tau$, defined from the energy density. 
This variable $\tilde{w}$ physically represents the ratio between the time $\tau$ and the equilibration time, which itself depends on $\tau$ through the temperature $T$. 
The pressure becomes isotropic in the limit $\tilde{w}\gg 1$, and is strongly anisotropic if $\tilde{w}\ll 1$. 
Thus, the larger $\eta/s$, the slower the quark distribution approaches  isotropy and chemical equilibrium~\cite{Du:2020dvp}.
We match the anisotropy parameter $\xi$ and the quark suppression factor $q_s$ to the results from QCD kinetic theory, and fix the late-time entropy density using the multiplicity measured by the ALICE collaboration in Pb--Pb collisions at $\sqrt{s_{\mathrm{NN}}}=5.02~\mathrm{TeV}$~\cite{ALICE:2016fbt}.

\begin{figure*}
    \centering
    \rotatebox{0}{
    \includegraphics[width=0.5\textwidth]{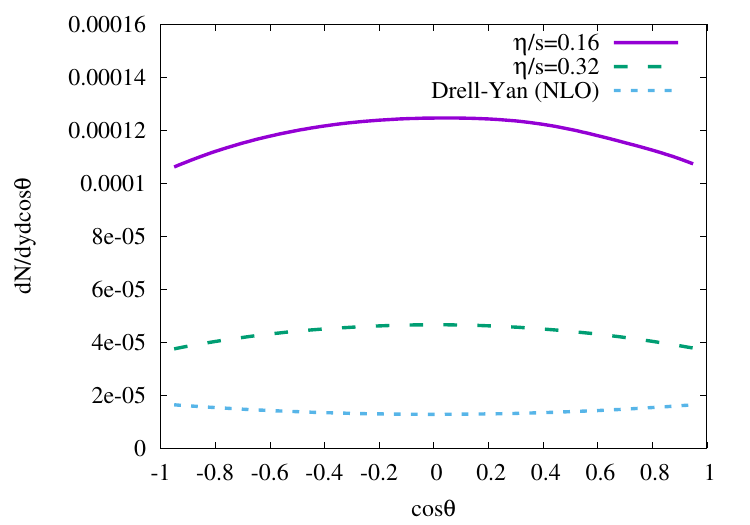}
    \includegraphics[width=0.5\textwidth]{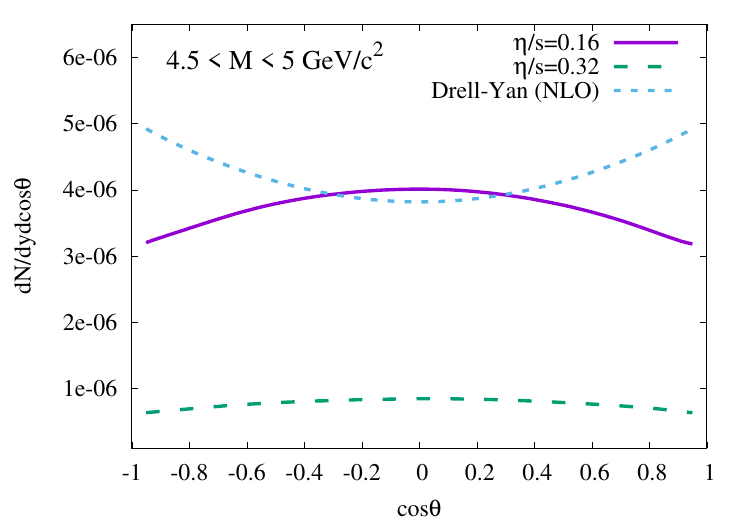}
    }
    \caption{Distribution of $\cos\theta$, defined by Eq.~(\ref{CS}), for quark-gluon plasma production with two values of the shear viscosity over entropy ratio $\eta/s$, and Drell-Yan production calculated at NLO, (left) in the $2.5 < M < 3 ~\mathrm{GeV}/c^2$ invariant mass bin and (right) in the $4.5 < M < 5~\mathrm{GeV}/c^2$.}
    \label{fig:CosTheta}
\end{figure*}
Results for the distribution of $\cos\theta$, defined by Eq.~(\ref{CS}), are displayed in Fig.~\ref{fig:CosTheta} for two ranges of invariant mass $M$, and two values of the parameter $\eta/s$, at midrapidity for $0-5\%$ central Pb--Pb collisions at $\sqrt{s_{\rm NN}}=5.02$~TeV. 
This distribution would be uniform if the distribution was isotropic. 
Instead, it is maximum for leptons emitted perpendicularly to the collision axis,  corresponding to $\cos\theta=0$. 
This is the expected consequence of the anisotropy of the quark momentum distribution, which is itself largest for purely transverse momenta. 
This maximum is more pronounced for the larger value of $M$: For $\eta/s=0.16$, the variation of the distribution as a function of $\cos\theta$ is $\sim 16 \%$ for the lower mass range $2.5 < M < 3~\mathrm{GeV}/c^2$ (left), and $\sim 30\%$ for the higher mass range  $4.5< M < 5~\mathrm{GeV}/c^2$. 
This is because higher masses are on average produced at earlier times, where the momentum anisotropy is larger. 
When the viscosity over entropy ratio $\eta/s$ increases, the number of produced dileptons decreases, due to the larger quark suppression~\cite{Coquet:2021lca}. 
The anisotropy also becomes more pronounced, because approach to equilibrium is slower. 
Simple scaling arguments show that pre-equilibrium effects scale with $\eta/s$ and $M$ like $(\eta/s)M^2$~\cite{Coquet:2021gms,Garcia-Montero:2023lrd}. 
Our calculation confirms this scaling.

\bigskip
{\it Drell-Yan dileptons. ---} 
At large invariant masses, dilepton production in hadron-hadron and nucleus-nucleus collisions at the LHC is dominated by the Drell-Yan process.
The leading-order production process is the same as for production in the quark-gluon plasma, namely, quark-antiquark annihilation. 
Therefore, the Drell-Yan process is usually considered as an experimentally irreducible background when studying production by a quark-gluon plasma.  
We will show that the angular distribution of leptons offers the possibility to disentangle the two processes. 

Drell-Yan production has been studied for several decades.  
The theory is much more advanced than that of production in the quark-gluon plasma. 
The process is calculated in perturbative QCD using collinear factorization.
The advantage over quark-gluon plasma production is that next-to-leading (NLO) calculations are available, which we use here. 

In addition to the leading-order quark-anti-quark annihilation $q \Bar{q} \rightarrow V \rightarrow l^+ l^-$, NLO processes include virtual corrections ($q \Bar{q} \rightarrow V g$), as well as Compton scattering ($q g \rightarrow V q$). These calculations are performed using the DrellYan Turbo software~\cite{Camarda:2019zyx}, with nuclear parton distributions (nPDFs) from EPPS~\cite{Eskola:2016oht}. 

For invariant masses $M\le 10~\mathrm{GeV}$, the perturbative calculation exhibits increasingly large uncertainties. These uncertainties mainly stem from the variation of the renormalization and factorization scales introduced in the collinear factorization approach. In this framework, the uncertainties are small at the $Z$-pole, but grow rapidly going towards the intermediate and low mass regions, as was shown in our previous publications, where the renormalization and factorization scales were varied by a factor of two~\cite{Coquet:2021lca}. For illustration purposes, these variations will not be considered in the following, and both scales are set equal to the dilepton mass. This does not alter the qualitative message of the present study.

Results for the $\cos\theta$ distribution of Drell-Yan dileptons are shown in Fig.~\ref{fig:CosTheta}. 
As expected, the angular distribution is opposite to that of quark-gluon plasma production, with a peak at $\cos\theta\ = \pm 1$ instead of $\cos\theta=0$. 
Note that the dependence on $\cos\theta$ is  weaker than the simple $1+\cos^2\theta$ expected from the leading-order Drell-Yan process. 
This is the effect of NLO corrections, and of quark momenta not being purely longitudinal. 
In the lower mass interval (left), production in the quark-gluon plasma dominates over Drell-Yan~\cite{Coquet:2021lca}. 
In the higher mass interval (right) it can be of similar magnitude or smaller, depending on the value of $\eta/s$.

%\section{Dilepton polarization}\label{sec:pola}

%For next-to-leading order production, the process $q \overline{q^{\prime}} \rightarrow l^+l^- + jet$ must be considered. In this case, the incoming partons will not be aligned anymore, as they must compensate for the momentum due to the produced jet. However, in the Collins-Soper frame, the $z$-axis still forms equal angles with the momenta of the initiating partons. This limits the impact of the jet momentum on the picture of this angle as the polar angle of the positive lepton, allowing the angular distribution can still be well approximated by the Born-level, which in the case of Drell-Yan is given by: $d \sigma / d \Omega \propto\left(1+\cos ^2 \theta\right)$.

\bigskip

{\it Total dilepton emission. ---} 

\begin{figure*}
    \centering
    \rotatebox{0}{
    \includegraphics[width=0.5\textwidth]{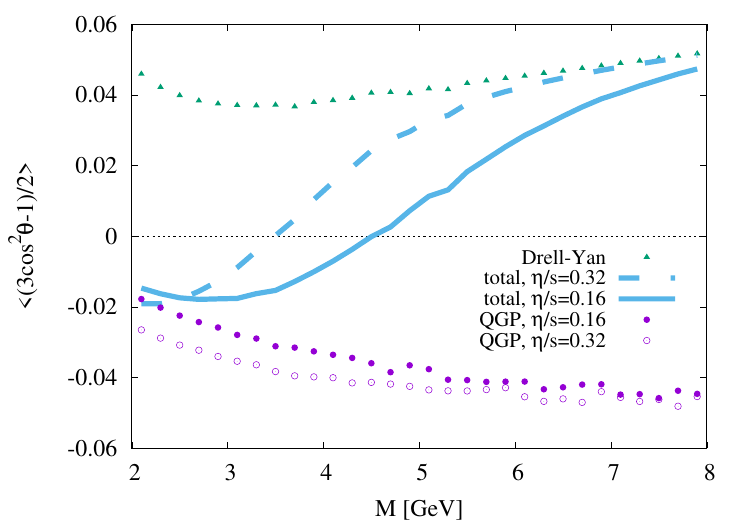}
    \includegraphics[width=0.5\textwidth]{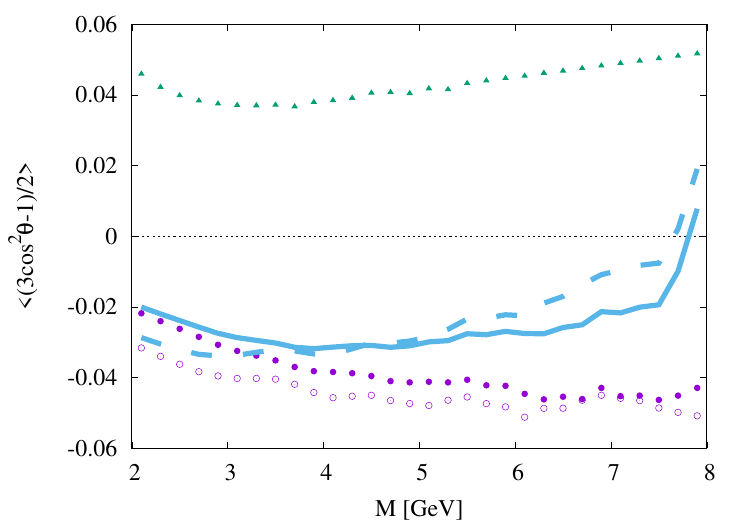}
    }
    \caption{Quadrupole moment of the angular distribution of leptons, defined as $\frac{1}{2}\left\langle 3\cos^2(\theta)-1\right\rangle$, for quark-gluon plasma production and Drell-Yan, as a function of the invariant mass $M$ of the dilepton pair. Left: default scenario with no quarks initially present. Right: assuming quark chemical equilibrium at all times.}
    \label{fig:MasseSpect}
\end{figure*}

We finally sum the contributions of quark-gluon plasma production and Drell-Yan, and study the dependence of the sum on $\cos\theta$. 
We choose as a measure of the angular asymmetry the average value of $(3\cos^2\theta-1)/2$, which corresponds to the quadrupole moment, and vanishes for an isotropic distribution: 
\begin{equation}
\label{asymmetry}
\left\langle\frac {3\cos^2\theta-1}{2}\right\rangle
\equiv
\frac{\int_{-1}^{1}d\cos\theta~\frac{1}{2}(3\cos^2\theta-1)\frac{dN}{d\cos\theta} }
{\int_{-1}^{1}d\cos\theta~\frac{dN}{d\cos\theta} }.
\end{equation}
Positive values of the quadrupole moment correspond to a prolate distribution and negative values to an oblate distribution. 
The extreme values are $1$, corresponding to all leptons emitted along the $z$ axis, and  $-\frac{1}{2}$, corresponding to all leptons emitted in the transverse plane. 
Note that if $dN/d\cos\theta\propto 1+\cos^2\theta$, which is the maximal angular dependence that one would expect from quark-antiquark annihilation, the quadrupole moment defined by Eq.~(\ref{asymmetry}) is only $\frac{1}{10}$. 
If the same distribution is rotated by an angle $\pi/2$, corresponding to emission by purely transverse quarks, the quadrupole moment is $-\frac{1}{20}$. 
One therefore expects the quadrupole moment to lie between $-0.05$ and $0.1$, that is, in a range ten times smaller than the mathematically-allowed interval between $-0.5$ and $1$. 
%\begin{figure}
%    \centering
%    \rotatebox{0}{
%    \includegraphics[width=0.5\textwidth]{anisotropy.pdf}
%    }
%    \caption{Quadrupole moment of the angular distribution of leptons, defined as $\frac{1}{2}\left\langle 3\cos^2(\theta)-1\right\rangle$, for quark-gluon plasma production and Drell-Yan, as a function of the invariant mass $M$ of the dilepton pair. Thick lines: default scenario with no quarks initially present. Thin lines: assuming quark chemical equilibrium at all times.}
%    \label{fig:MasseSpect}
%\end{figure}

Fig.~\ref{fig:MasseSpect} displays the variation of the quadrupole moment  with the invariant mass $M$ for both processes individually, as well as for the sum. 
For quark-gluon plasma production, the quadrupole moment is negative as expected, corresponding to an oblate distribution. 
It goes to zero for small values of $M$, which are produced at late times when the pressure is isotropic.\footnote{Note, however, that our calculation, which neglects transverse flow, is not reliable for small $M$~\cite{Speranza:2018osi}, typically below 2~GeV.} 
As $M$ increases, it rapidly approaches the lower bound $-0.05$. 
The variation with $\eta/s$ scales with $(\eta/s)M^2$, as expected by dimensional analysis~\cite{Coquet:2021gms}. 
If quarks are in chemical equilibrium (Fig.\ref{fig:MasseSpect} right), the asymptotic limit $-0.05$ is approached somewhat faster. 

For Drell-Yan production, the quadrupole moment is positive, corresponding to a prolate distribution, and varies weakly with $M$. 
Note that it is smaller by a factor $\sim 2$ than the upper bound $\frac{1}{10}$. 

If one sums the contributions of quark-gluon plasma production and Drell-Yan before evaluating the asymmetry, one finds that the asymmetry is negative for the lower values of $M$, where Drell-Yan is negligible, and becomes positive for larger values of $M$, where Drell-Yan dominates. 
The value of $M$ for which the transition occurs depends on $\eta/s$. 
The smaller $\eta/s$, the more copiously the quark-gluon plasma produces lepton pairs, and negative asymmetries are observed up to $M\sim 4$~GeV. 
If quarks are in chemical equilibrum (Fig.\ref{fig:MasseSpect} right), dilepton production in the quark-gluon plasma dominates over Drell-Yan production up to $M\sim 7-8$~GeV~\cite{Coquet:2021lca}, so that the asymmetry remains negative.

%\section{Conclusion}
In summary, we have introduced a new observable, the quadrupole moment of the angular distribution of leptons, which is negative for dileptons produced in the quark-gluon plasma, and positive for Drell-Yan dileptons. 
Measurement of this observable in future LHC experiments will provide the first experimental proof that the pressure tensor of the quark-gluon plasma is oblate at early times, and provide a first direct experimental  constraint on the isotropization time of the quark-gluon plasma.

\begin{acknowledgments}
%This work is supported in part by the Deutsche Forschungsgemeinschaft (DFG, German Research Foundation) through the CRC-TR 211 ``Strong-interaction matter under extreme conditions'' project number 315477589 – TRR 211 and in part in the framework of the GLUODYNAMICS project funded by the ``P2IO LabEx (ANR-10-LABX-0038)'' in the framework ``Investissements d'Avenir'' (ANR-11-IDEX-0003-01) managed by the Agence Nationale de la Recherche (ANR), France.
This work is supported in part by the Deutsche Forschungsgemeinschaft (DFG, German Research Foundation) through the CRC-TR 211 ``Strong-interaction matter under extreme conditions'' project number 315477589 – TRR 211, and in part in the framework of the GLUODYNAMICS project funded by the ``P2IO LabEx (ANR-10-LABX-0038)'' in the framework ``Investissements d'Avenir'' (ANR-11-IDEX-0003-01) managed by the Agence Nationale de la Recherche (ANR), France, and in part by Xunta de Galicia (Centro singular de investigacion de
Galicia accreditation 2019-2022), European Union ERDF,
the “Maria de Maeztu” Units of Excellence program under
project CEX2020-001035-M, the Spanish Research State
Agency under project PID2020-119632GB-I00, and European Research Council under project ERC-2018-ADG835105 YoctoLHC.
\end{acknowledgments}

%\bibliography{main}% Produces the bibliography via BibTeX.

\begin{thebibliography}{99}

%\cite{Busza:2018rrf}
\bibitem{Busza:2018rrf}
W.~Busza, K.~Rajagopal and W.~van der Schee,
%``Heavy Ion Collisions: The Big Picture, and the Big Questions,''
Ann. Rev. Nucl. Part. Sci. \textbf{68}, 339-376 (2018)
doi:10.1146/annurev-nucl-101917-020852
[arXiv:1802.04801 [hep-ph]].
%636 citations counted in INSPIRE as of 15 Apr 2024

%\cite{Schlichting:2019abc}
\bibitem{Schlichting:2019abc}
S.~Schlichting and D.~Teaney,
%``The First fm/c of Heavy-Ion Collisions,''
Ann. Rev. Nucl. Part. Sci. \textbf{69}, 447-476 (2019)
doi:10.1146/annurev-nucl-101918-023825
[arXiv:1908.02113 [nucl-th]].
%92 citations counted in INSPIRE as of 03 Apr 2024

%\cite{Berges:2020fwq}
\bibitem{Berges:2020fwq}
J.~Berges, M.~P.~Heller, A.~Mazeliauskas and R.~Venugopalan,
%``QCD thermalization: Ab initio approaches and interdisciplinary connections,''
Rev. Mod. Phys. \textbf{93}, no.3, 035003 (2021)
doi:10.1103/RevModPhys.93.035003
[arXiv:2005.12299 [hep-th]].
%199 citations counted in INSPIRE as of 12 Apr 2024

%\cite{Bjorken:1982qr}
\bibitem{Bjorken:1982qr}
J.~D.~Bjorken,
%``Highly Relativistic Nucleus-Nucleus Collisions: The Central Rapidity Region,''
Phys. Rev. D \textbf{27}, 140-151 (1983)
doi:10.1103/PhysRevD.27.140
%3634 citations counted in INSPIRE as of 11 Apr 2024

%\cite{Baier:2000sb}
\bibitem{Baier:2000sb}
R.~Baier, A.~H.~Mueller, D.~Schiff and D.~T.~Son,
%``'Bottom up' thermalization in heavy ion collisions,''
Phys. Lett. B \textbf{502}, 51-58 (2001)
doi:10.1016/S0370-2693(01)00191-5
[arXiv:hep-ph/0009237 [hep-ph]].
%662 citations counted in INSPIRE as of 12 Apr 2024

%\cite{Epelbaum:2013ekf}
\bibitem{Epelbaum:2013ekf}
T.~Epelbaum and F.~Gelis,
%``Pressure isotropization in high energy heavy ion collisions,''
Phys. Rev. Lett. \textbf{111}, 232301 (2013)
doi:10.1103/PhysRevLett.111.232301
[arXiv:1307.2214 [hep-ph]].
%205 citations counted in INSPIRE as of 20 Mar 2024

%\cite{Berges:2013fga}
\bibitem{Berges:2013fga}
J.~Berges, K.~Boguslavski, S.~Schlichting and R.~Venugopalan,
%``Universal attractor in a highly occupied non-Abelian plasma,''
Phys. Rev. D \textbf{89}, no.11, 114007 (2014)
doi:10.1103/PhysRevD.89.114007
[arXiv:1311.3005 [hep-ph]].
%229 citations counted in INSPIRE as of 20 Mar 2024

%\cite{Kurkela:2015qoa}
\bibitem{Kurkela:2015qoa}
A.~Kurkela and Y.~Zhu,
%``Isotropization and hydrodynamization in weakly coupled heavy-ion collisions,''
Phys. Rev. Lett. \textbf{115}, no.18, 182301 (2015)
doi:10.1103/PhysRevLett.115.182301
[arXiv:1506.06647 [hep-ph]].
%252 citations counted in INSPIRE as of 12 Apr 2024

%\cite{Kurkela:2018vqr}
\bibitem{Kurkela:2018vqr}
A.~Kurkela, A.~Mazeliauskas, J.~F.~Paquet, S.~Schlichting and D.~Teaney,
%``Effective kinetic description of event-by-event pre-equilibrium dynamics in high-energy heavy-ion collisions,''
Phys. Rev. C \textbf{99}, no.3, 034910 (2019)
doi:10.1103/PhysRevC.99.034910
[arXiv:1805.00961 [hep-ph]].
%162 citations counted in INSPIRE as of 09 Apr 2024

%\cite{Alqahtani:2017mhy}
\bibitem{Alqahtani:2017mhy}
M.~Alqahtani, M.~Nopoush and M.~Strickland,
%``Relativistic anisotropic hydrodynamics,''
Prog. Part. Nucl. Phys. \textbf{101}, 204-248 (2018)
doi:10.1016/j.ppnp.2018.05.004
[arXiv:1712.03282 [nucl-th]].
%116 citations counted in INSPIRE as of 15 Apr 2024

%\cite{Hoyer:1986pp}
\bibitem{Hoyer:1986pp}
P.~Hoyer,
%``Particle Polarization as a Signal of Plasma Formation,''
Phys. Lett. B \textbf{187}, 162-164 (1987)
doi:10.1016/0370-2693(87)90091-8
%21 citations counted in INSPIRE as of 29 Feb 2024

%\cite{Bratkovskaya:1995kh}
\bibitem{Bratkovskaya:1995kh}
E.~L.~Bratkovskaya, O.~V.~Teryaev and V.~D.~Toneev,
%``Anisotropy of dilepton emission from nuclear collisions,''
Phys. Lett. B \textbf{348}, 283-289 (1995)
doi:10.1016/0370-2693(95)00164-G
%59 citations counted in INSPIRE as of 29 Feb 2024

%\cite{Ipp:2007ng}
\bibitem{Ipp:2007ng}
A.~Ipp, A.~Di Piazza, J.~Evers and C.~H.~Keitel,
%``Photon polarization as a probe for quark-gluon plasma dynamics,''
Phys. Lett. B \textbf{666}, 315-319 (2008)
doi:10.1016/j.physletb.2008.07.076
[arXiv:0710.5700 [hep-ph]].
%36 citations counted in INSPIRE as of 05 Mar 2024

%\cite{Baym:2014qfa}
\bibitem{Baym:2014qfa}
G.~Baym and T.~Hatsuda,
%``Polarization of Direct Photons from Gluon Anisotropy in Ultrarelativistic Heavy Ion Collisions,''
PTEP \textbf{2015}, no.3, 031D01 (2015)
doi:10.1093/ptep/ptv024
[arXiv:1405.1376 [nucl-th]].
%10 citations counted in INSPIRE as of 05 Mar 2024

%\cite{Drell:1970wh}
\bibitem{Drell:1970wh}
S.~D.~Drell and T.~M.~Yan,
%``Massive Lepton Pair Production in Hadron-Hadron Collisions at High-Energies,''
Phys. Rev. Lett. \textbf{25}, 316-320 (1970)
[erratum: Phys. Rev. Lett. \textbf{25}, 902 (1970)]
doi:10.1103/PhysRevLett.25.316
%1835 citations counted in INSPIRE as of 05 Apr 2024

%\cite{McLerran:1984ay}
\bibitem{McLerran:1984ay}
L.~D.~McLerran and T.~Toimela,
%``Photon and Dilepton Emission from the Quark - Gluon Plasma: Some General Considerations,''
Phys. Rev. D \textbf{31}, 545 (1985)
doi:10.1103/PhysRevD.31.545
%586 citations counted in INSPIRE as of 03 Apr 2024

%\cite{Coquet:2021lca}
\bibitem{Coquet:2021lca}
M.~Coquet, X.~Du, J.~Y.~Ollitrault, S.~Schlichting and M.~Winn,
%``Intermediate mass dileptons as pre-equilibrium probes in heavy ion collisions,''
Phys. Lett. B \textbf{821}, 136626 (2021)
doi:10.1016/j.physletb.2021.136626
[arXiv:2104.07622 [nucl-th]].
%30 citations counted in INSPIRE as of 12 Apr 2024

%\cite{Rapp:2013nxa}
\bibitem{Rapp:2013nxa}
R.~Rapp,
%``Dilepton Spectroscopy of QCD Matter at Collider Energies,''
Adv. High Energy Phys. \textbf{2013}, 148253 (2013)
doi:10.1155/2013/148253
[arXiv:1304.2309 [hep-ph]].
%139 citations counted in INSPIRE as of 03 Apr 2024

%\cite{Song:2018xca}
\bibitem{Song:2018xca}
T.~Song, W.~Cassing, P.~Moreau and E.~Bratkovskaya,
%``Open charm and dileptons from relativistic heavy-ion collisions,''
Phys. Rev. C \textbf{97}, no.6, 064907 (2018)
doi:10.1103/PhysRevC.97.064907
[arXiv:1803.02698 [nucl-th]].
%41 citations counted in INSPIRE as of 03 Apr 2024

%\cite{Andronic:2015wma}
\bibitem{Andronic:2015wma}
A.~Andronic, F.~Arleo, R.~Arnaldi, A.~Beraudo, E.~Bruna, D.~Caffarri, Z.~C.~del Valle, J.~G.~Contreras, T.~Dahms and A.~Dainese, \textit{et al.}
%``Heavy-flavour and quarkonium production in the LHC era: from proton\textendash{}proton to heavy-ion collisions,''
Eur. Phys. J. C \textbf{76}, no.3, 107 (2016)
doi:10.1140/epjc/s10052-015-3819-5
[arXiv:1506.03981 [nucl-ex]].
%660 citations counted in INSPIRE as of 10 Apr 2024

%\cite{ALICE:2018ael}
\bibitem{ALICE:2018ael}
S.~Acharya \textit{et al.} [ALICE],
%``Measurement of dielectron production in central Pb-Pb collisions at $\sqrt{{\textit{s}}_{\mathrm{NN}}}$ = 2.76 TeV,''
Phys. Rev. C \textbf{99}, no.2, 024002 (2019)
doi:10.1103/PhysRevC.99.024002
[arXiv:1807.00923 [nucl-ex]].
%43 citations counted in INSPIRE as of 26 Mar 2024

%\cite{ALICE:2023jef}
\bibitem{ALICE:2023jef}
S.~Acharya \textit{et al.} [ALICE],
%``Dielectron production in central Pb$-$Pb collisions at $\sqrt{s_\mathrm{NN}}$ = 5.02 TeV,''
[arXiv:2308.16704 [nucl-ex]].
%4 citations counted in INSPIRE as of 09 Apr 2024

%\cite{LHCb:2018roe}
\bibitem{LHCb:2018roe}
R.~Aaij \textit{et al.} [LHCb],
%``Physics case for an LHCb Upgrade II - Opportunities in flavour physics, and beyond, in the HL-LHC era,''
[arXiv:1808.08865 [hep-ex]].
%453 citations counted in INSPIRE as of 15 Apr 2024

%\cite{Adamova:2019vkf}
\bibitem{Adamova:2019vkf}
D.~Adamov\'a, G.~Aglieri Rinella, M.~Agnello, Z.~Ahammed, D.~Aleksandrov, A.~Alici, A.~Alkin, T.~Alt, I.~Altsybeev and D.~Andreou, \textit{et al.}
%``A next-generation LHC heavy-ion experiment,''
[arXiv:1902.01211 [physics.ins-det]].
%53 citations counted in INSPIRE as of 28 Mar 2024

%\cite{Baym:2017qxy}
\bibitem{Baym:2017qxy}
G.~Baym, T.~Hatsuda and M.~Strickland,
%``Virtual photon polarization in ultrarelativistic heavy-ion collisions,''
Phys. Rev. C \textbf{95}, no.4, 044907 (2017)
doi:10.1103/PhysRevC.95.044907
[arXiv:1702.05906 [nucl-th]].
%20 citations counted in INSPIRE as of 29 Feb 2024

%\cite{Speranza:2018osi}
\bibitem{Speranza:2018osi}
E.~Speranza, A.~Jaiswal and B.~Friman,
%``Virtual photon polarization and dilepton anisotropy in relativistic nucleus\textendash{}nucleus collisions,''
Phys. Lett. B \textbf{782}, 395-400 (2018)
doi:10.1016/j.physletb.2018.05.053
[arXiv:1802.02479 [hep-ph]].
%17 citations counted in INSPIRE as of 29 Feb 2024

%\cite{Collins:1977iv}
\bibitem{Collins:1977iv}
J.~C.~Collins and D.~E.~Soper,
%``Angular Distribution of Dileptons in High-Energy Hadron Collisions,''
Phys. Rev. D \textbf{16}, 2219 (1977)
doi:10.1103/PhysRevD.16.2219
%1082 citations counted in INSPIRE as of 12 Apr 2024

%\cite{NA60:2008iqj}
\bibitem{NA60:2008iqj}
R.~Arnaldi \textit{et al.} [NA60],
%``First results on angular distributions of thermal dileptons in nuclear collisions,''
Phys. Rev. Lett. \textbf{102}, 222301 (2009)
doi:10.1103/PhysRevLett.102.222301
[arXiv:0812.3100 [nucl-ex]].
%34 citations counted in INSPIRE as of 11 Mar 2024

%\cite{Coquet:2021gms}
\bibitem{Coquet:2021gms}
M.~Coquet, X.~Du, J.~Y.~Ollitrault, S.~Schlichting and M.~Winn,
%``Transverse mass scaling of dilepton radiation off a quark-gluon plasma,''
Nucl. Phys. A \textbf{1030}, 122579 (2023)
doi:10.1016/j.nuclphysa.2022.122579
[arXiv:2112.13876 [nucl-th]].
%12 citations counted in INSPIRE as of 03 Apr 2024

%\cite{Schwartz:2014sze}
\bibitem{Schwartz:2014sze}
M.~D.~Schwartz,
%``Quantum Field Theory and the Standard Model,''
Cambridge University Press, 2014,
ISBN 978-1-107-03473-0, 978-1-107-03473-0
%159 citations counted in INSPIRE as of 15 Apr 2024

%\cite{Romatschke:2003ms}
\bibitem{Romatschke:2003ms}
P.~Romatschke and M.~Strickland,
%``Collective modes of an anisotropic quark gluon plasma,''
Phys. Rev. D \textbf{68}, 036004 (2003)
doi:10.1103/PhysRevD.68.036004
[arXiv:hep-ph/0304092 [hep-ph]].
%492 citations counted in INSPIRE as of 03 Apr 2024

%\cite{Martinez:2007pjh}
\bibitem{Martinez:2007pjh}
M.~Martinez and M.~Strickland,
%``Measuring QGP thermalization time with dileptons,''
Phys. Rev. Lett. \textbf{100}, 102301 (2008)
doi:10.1103/PhysRevLett.100.102301
[arXiv:0709.3576 [hep-ph]].
%76 citations counted in INSPIRE as of 11 Mar 2024

%\cite{Lappi:2006fp}
\bibitem{Lappi:2006fp}
T.~Lappi and L.~McLerran,
%``Some features of the glasma,''
Nucl. Phys. A \textbf{772}, 200-212 (2006)
doi:10.1016/j.nuclphysa.2006.04.001
[arXiv:hep-ph/0602189 [hep-ph]].
%568 citations counted in INSPIRE as of 09 Apr 2024

%\cite{Kurkela:2018oqw}
\bibitem{Kurkela:2018oqw}
A.~Kurkela and A.~Mazeliauskas,
%``Chemical equilibration in weakly coupled QCD,''
Phys. Rev. D \textbf{99}, no.5, 054018 (2019)
doi:10.1103/PhysRevD.99.054018
[arXiv:1811.03068 [hep-ph]].
%73 citations counted in INSPIRE as of 12 Apr 2024

%\cite{Kurkela:2018xxd}
\bibitem{Kurkela:2018xxd}
A.~Kurkela and A.~Mazeliauskas,
%``Chemical Equilibration in Hadronic Collisions,''
Phys. Rev. Lett. \textbf{122}, 142301 (2019)
doi:10.1103/PhysRevLett.122.142301
[arXiv:1811.03040 [hep-ph]].
%67 citations counted in INSPIRE as of 12 Apr 2024

%\cite{Vredevoogd:2008id}
\bibitem{Vredevoogd:2008id}
J.~Vredevoogd and S.~Pratt,
%``Universal Flow in the First Stage of Relativistic Heavy Ion Collisions,''
Phys. Rev. C \textbf{79}, 044915 (2009)
doi:10.1103/PhysRevC.79.044915
[arXiv:0810.4325 [nucl-th]].
%107 citations counted in INSPIRE as of 09 Apr 2024

%\cite{Giacalone:2019ldn}
\bibitem{Giacalone:2019ldn}
G.~Giacalone, A.~Mazeliauskas and S.~Schlichting,
%``Hydrodynamic attractors, initial state energy and particle production in relativistic nuclear collisions,''
Phys. Rev. Lett. \textbf{123}, no.26, 262301 (2019)
doi:10.1103/PhysRevLett.123.262301
[arXiv:1908.02866 [hep-ph]].
%84 citations counted in INSPIRE as of 12 Apr 2024

%\cite{Du:2020dvp}
\bibitem{Du:2020dvp}
X.~Du and S.~Schlichting,
%``Equilibration of weakly coupled QCD plasmas,''
Phys. Rev. D \textbf{104}, no.5, 054011 (2021)
doi:10.1103/PhysRevD.104.054011
[arXiv:2012.09079 [hep-ph]].
%46 citations counted in INSPIRE as of 12 Apr 2024

%\cite{Du:2020zqg}
\bibitem{Du:2020zqg}
X.~Du and S.~Schlichting,
%``Equilibration of the Quark-Gluon Plasma at Finite Net-Baryon Density in QCD Kinetic Theory,''
Phys. Rev. Lett. \textbf{127}, no.12, 122301 (2021)
doi:10.1103/PhysRevLett.127.122301
[arXiv:2012.09068 [hep-ph]].
%40 citations counted in INSPIRE as of 12 Apr 2024

%\cite{Kovtun:2004de}
\bibitem{Kovtun:2004de}
P.~Kovtun, D.~T.~Son and A.~O.~Starinets,
%``Viscosity in strongly interacting quantum field theories from black hole physics,''
Phys. Rev. Lett. \textbf{94}, 111601 (2005)
doi:10.1103/PhysRevLett.94.111601
[arXiv:hep-th/0405231 [hep-th]].
%2897 citations counted in INSPIRE as of 15 Apr 2024

%\cite{Romatschke:2007mq}
\bibitem{Romatschke:2007mq}
P.~Romatschke and U.~Romatschke,
%``Viscosity Information from Relativistic Nuclear Collisions: How Perfect is the Fluid Observed at RHIC?,''
Phys. Rev. Lett. \textbf{99}, 172301 (2007)
doi:10.1103/PhysRevLett.99.172301
[arXiv:0706.1522 [nucl-th]].
%1168 citations counted in INSPIRE as of 08 Apr 2024

%\cite{Heinz:2013th}
\bibitem{Heinz:2013th}
U.~Heinz and R.~Snellings,
%``Collective flow and viscosity in relativistic heavy-ion collisions,''
Ann. Rev. Nucl. Part. Sci. \textbf{63}, 123-151 (2013)
doi:10.1146/annurev-nucl-102212-170540
[arXiv:1301.2826 [nucl-th]].
%1282 citations counted in INSPIRE as of 15 Apr 2024

%\cite{Heller:2016rtz}
\bibitem{Heller:2016rtz}
M.~P.~Heller, A.~Kurkela, M.~Spali\'nski and V.~Svensson,
%``Hydrodynamization in kinetic theory: Transient modes and the gradient expansion,''
Phys. Rev. D \textbf{97}, no.9, 091503 (2018)
doi:10.1103/PhysRevD.97.091503
[arXiv:1609.04803 [nucl-th]].
%164 citations counted in INSPIRE as of 03 Apr 2024

%\cite{ALICE:2016fbt}
\bibitem{ALICE:2016fbt}
J.~Adam \textit{et al.} [ALICE],
%``Centrality dependence of the pseudorapidity density distribution for charged particles in Pb-Pb collisions at $\sqrt{s_{\rm NN}}=5.02$ TeV,''
Phys. Lett. B \textbf{772}, 567-577 (2017)
doi:10.1016/j.physletb.2017.07.017
[arXiv:1612.08966 [nucl-ex]].
%130 citations counted in INSPIRE as of 12 Apr 2024

%\cite{Garcia-Montero:2023lrd}
\bibitem{Garcia-Montero:2023lrd}
O.~Garcia-Montero, A.~Mazeliauskas, P.~Plaschke and S.~Schlichting,
%``Pre-equilibrium photons from the early stages of heavy-ion collisions,''
JHEP \textbf{03}, 053 (2024)
doi:10.1007/JHEP03(2024)053
[arXiv:2308.09747 [hep-ph]].
%9 citations counted in INSPIRE as of 09 Apr 2024

%\cite{Camarda:2019zyx}
\bibitem{Camarda:2019zyx}
S.~Camarda, M.~Boonekamp, G.~Bozzi, S.~Catani, L.~Cieri, J.~Cuth, G.~Ferrera, D.~de Florian, A.~Glazov and M.~Grazzini, \textit{et al.}
%``DYTurbo: Fast predictions for Drell-Yan processes,''
Eur. Phys. J. C \textbf{80}, no.3, 251 (2020)
[erratum: Eur. Phys. J. C \textbf{80}, no.5, 440 (2020)]
doi:10.1140/epjc/s10052-020-7757-5
[arXiv:1910.07049 [hep-ph]].
%89 citations counted in INSPIRE as of 12 Apr 2024

%\cite{Eskola:2016oht}
\bibitem{Eskola:2016oht}
K.~J.~Eskola, P.~Paakkinen, H.~Paukkunen and C.~A.~Salgado,
%``EPPS16: Nuclear parton distributions with LHC data,''
Eur. Phys. J. C \textbf{77}, no.3, 163 (2017)
doi:10.1140/epjc/s10052-017-4725-9
[arXiv:1612.05741 [hep-ph]].
%617 citations counted in INSPIRE as of 15 Apr 2024

\end{thebibliography}

\end{document}